\documentclass[journal=jacsat,manuscript=article]{achemso}
 \usepackage{chemformula} 
 \usepackage[T1]{fontenc}
 \usepackage{float}
 \usepackage[labelfont=bf,labelsep=period]{caption}
 \usepackage{sectsty}
 \usepackage[utf8]{inputenc}
 \usepackage{array}
 \usepackage{geometry}
 \usepackage{setspace}
 \usepackage{xkeyval}
 \usepackage{xcolor}
 \usepackage{natbib}
 
\author{Zixiang Wei}
\affiliation{Department of Chemical and Materials Engineering, University of Alberta, Alberta T6G 1H9, Canada}
\author{Miaosi Li}
\affiliation{School of Engineering, RMIT University, Melbourne, Victoria 3001, Australia}
\email{miaosi.li@rmit.edu.au}
\author{Hongbo Zeng}
\affiliation{Department of Chemical and Materials Engineering, University of Alberta, Alberta T6G 1H9, Canada}
\author{Xuehua Zhang}
\affiliation{Department of Chemical and Materials Engineering, University of Alberta, Alberta T6G 1H9, Canada}
\title {Integrated nanoextraction and colorimetric reactions in surface nanodroplets for combinative analysis}
\keywords{femtoliter droplets, nanoextraction, multicomponent matrix, colorimetric reaction,  anti-counterfeiting}

\begin{document}
\begin{abstract}

A combinative approach for chemical analysis makes it possible to distinguish a mixture of a large number of compounds from other mixtures in a single step.
This work demonstrates a combinative analysis approach by using surface nanodroplets for integrating nanoextraction and colorimetric reactions for the identification of multicomponent mixtures. 
The model analytes are acidic compounds dissolved in oil that are extracted into aqueous droplets on a solid substrate. The proton from acid dissociation reacts with the halochromic chemical compounds inside the droplets, leading to the color change of the droplets. The rate of the colorimetric reaction exhibits certain specificity for the acid type, distinguishing acid mixtures with the same pH value. The underlying principle is that the acid transport rate is associated with the partition coefficient and the dissociation constant of the acid, in addition to the concentration in oil.  As a demonstration, we showed that droplet-based combinative analysis can be applied for anti-counterfeiting of various alcoholic spirits by comparing decolor time of organic acid mixtures in the spirits. The readout can be done by using a common hand-hold mobile phone. 

 \end{abstract}

\section{Introduction}

Surface nanodroplets are liquid droplets on a solid substrate in contact with an immiscible fluid medium. These droplets are tens to several hundreds nanometer in the apex and several hundred nanometer to tens of micrometer in the base diameter.\cite{zhang2012transient} These droplets with femtoliter or even less in the volume are stable for a long time against evaporation or dissolution, possessing desirable characteristics that enable novel applications in the field of chemical analysis. As compounds with higher solubility in the droplet liquid can be extracted and concentrated in surface nanodroplets,  liquid-liquid nanoextraction can be employed for in-situ ultrafast analysis of liquid samples with very small volume (such as sneezing droplets) \cite{ll2}, and for automated determination of partition coefficient(LgP) of drug molecules in a continuous flow system.\cite{snd2} 
Thanks to the large surface-to-volume ratio, nanodroplets also provide a steady and well-controlled microcompartmentalized environment for biphasic reactions in need of reagents in the immiscible phases of droplets and an external solution in the flow. \cite{pcsf3} Many chemical reactions in microdroplets are found to be faster and more efficient than their macroscopic counterparts in a bulk medium\cite{react}. Even some reactions impossible in the bulk system without catalysts can occur spontaneously in microdroplets. Combining reactive component and non-reactive extractant in surface nanodroplets is especially powerful, as these binary droplets provide enhanced sensitivity from a high concentration of the analyte extracted into the droplets. The concept of integrating nanoextraction and chemical detection has been demonstrated in quantitative in-situ detection by surface-enhanced Raman spectroscopy\cite{snd2}.
 
Recent advance in nanodroplet generation by solvent exchange offers new opportunities for nanodroplet-based chemical analysis, as it is possible to produce multicomponent droplets by this simple method.\cite{li2018formation} The ratio of different components in the droplets is controlled by their respective oversaturation created by the solvent exchange.\cite{li2018formation} Droplets can contain various types of solvents, their combinations or solid solutes, \cite{snd3,qian2019} providing a general plat form for droplet-based sensing. Up to now, what rarely explored is the characteristic time scale from integrated nanoextraction and chemical reactions in surface nanodroplets for chemical identification. For a given external concentration of the compound, the extraction rate of surface nanodroplets is not specific to the compound but is associated with its solubility difference in the droplet and in the surrounding liquid at equilibrium (i.e. partition coefficient). If the extracted analyte reacts with a chemical in the multicomponent droplets, the reaction rate will reflect the solubility of the analyte, allowing for categorizing compounds based on the solubility. If other property of the analyte can also play a role in the reaction rate, the timescale of the product from the droplet reaction will be resulted from the synergistic effect from the solubility and other property, and hence more specific to the compound. With increasing the number of analytes extracted to the reactive droplets, the time scale for the product can be more complex from combined effects of all components in the solubility family, and be even unique for the mixture. To demonstrate this concept of combinative analysis by using reactive surface nanodroplets, we choose the detection of a mixture of organic acids in one-step.  

Detection of acidic compounds is critically important in many areas, such as controlling the flavors and quality of food, and anti-corrosion of petroleum and gasoline products. \cite{OA1,OA2,OA3,OA4,OA6}. In many cases, acidic compounds are presented in the oil phase, ranging from heavy oils, petroleum products to cooking oil, beverages and essential oils for well being and medical use \cite{OA1,OA2,OA3}.  
Colorimetric detection of organic acids in an aqueous medium could contribute to eliminating the risks of using highly toxic organic solvents in analysis of acids in oil. However, there are challenges for detecting acidic compounds in oil by using the aqueous solution of the indicator, as the reaction requires access to the reagents in two immiscible liquids, oil and water. Intrinsic characteristics of biphasic reactions could be severely influenced by the inefficient mixing, due to the limited interface between oil and water in the bulk medium. Thus it often requires a long time for acidic compounds to achieve equilibrium within the oil-water system.\cite{pcsf4}
Furthermore, as the colorimetric reaction occurred in aqueous solution reveals the acids that are transferred into the water, the acid concentration in the oil phase may be determined only when the oil-water partition coefficient (lgP) of the acid is known. \cite{LogP1,LogP2} In many cases, lgP of the acid is not necessarily available for given detection conditions and the determination of lgP is also time and equipment consuming. \cite{pcsf1,pcsf2,pcsf3,lgpx} To overcome the above challenges, a detection process that can reduce the time for partition equilibrium and can minimize the chemical consumption becomes highly valuable for detection of acidic compounds in oil.  

In this work, we will show a nanodroplet-based combinative approach for colorimetric detection and identification of organic acid family in oils. The droplets containing pH indicators show visible and sensitive colorimetric response to organic acids extracted from the surrounding oil flow. The rate of the response is related to the type of acids, due to the effect of partition coefficient and dissociation constant of the acid on the transport into the droplets.  
The colorimetric reactions in the droplets are utilized to identify the profiles of acidic compounds in a mixture. The potential application of combinative analysis by nanodroplets is demonstrated in anti-counterfeiting of expensive alcohol drinks.

\section{Experimental section}

\subsection{Chemicals and materials} 
 1-octanol (95\%, Fisher Scientific, Canada) was selected as a modeling oil because the partition coefficients of many chemicals in octanol and water are available in the literature. The halochromic compounds were Bromocresol green (BG, yellow to blue at pH 3.8 to 5.4), bromocresol purple (BP, yellow to purple at pH 5.2 to 6.8). All halochromic  compounds were purchased from Fisher Scientific (Canada). Several different types of acids in oils were tested, including citric acid (99.5\%), gallic acid (97.5\%), 4-hydroxybenozic acid (99\%), benzoic acid (99.5\%) and trimesic acid (99\%), all purchased from Sigma-Alderich (Canada). Acetic acid (99.7\%) and heptanoic acid (98\%) were purchased from Fisher Scientific (Canada). The LgP and dissociation constant (pKa) of each acid are shown in Table \ref{lgP}. Water was obtained from Milli-Q water purification unit (Millipore Corporation, Boston, MA, USA).

\subsection{Formation of pH sensitive femtoliter droplets}
\begin{figure}[htp]
	\includegraphics[trim={0cm 0cm 0cm 0cm}, clip, width=0.95\columnwidth]{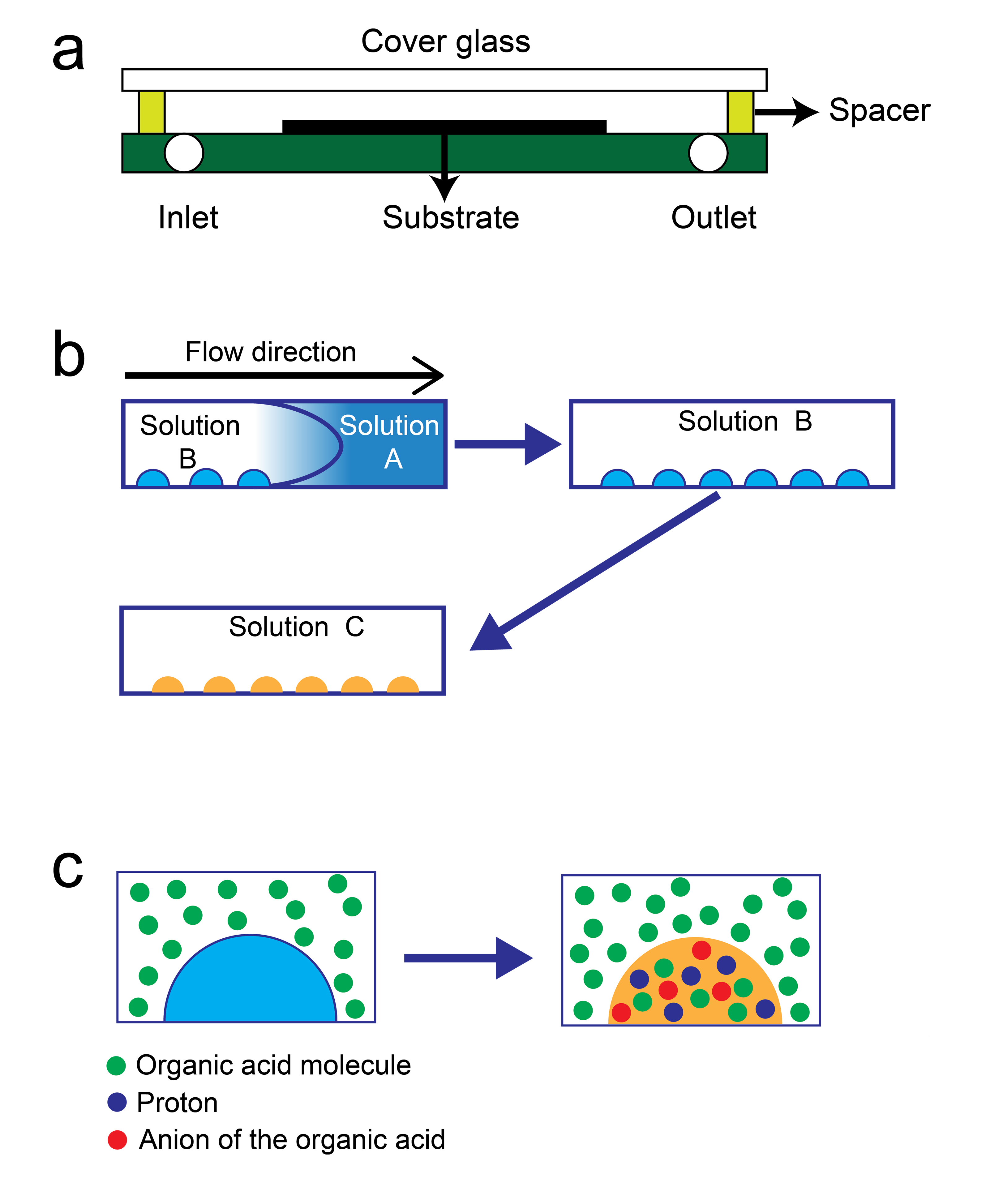}
	\caption{Schematic of preparing and applying surface droplets for the extraction and detection of acidic compounds in oil. (a) The sketch of the fluid chamber.(b) The formation process of the surface droplets containing pH indicator by solvent exchange and detection of the acid in oil (Solution C) by droplets contacting with the Oil flow.(c) Acid molecules in the oil phase are extracted into the aqueous droplets and react with halochromic compound in the droplets. }
	\label{f1}
\end{figure}

Formation of surface droplets was performed in a fluid chamber through the procedure of solvent exchange \cite{snd3}. The setup of the fluid chamber is similar to the one used in our previous report \cite{miaosi2018}, as shown in \ref{f1}a. In brief, silicon substrate (University Wafer, USA) was coated by 3-aminopropyl triethoxysilane (APTES) \cite{nanoroughness}. The APTES-coated substrate was cut into a slab and placed on the bottom of the chamber. The top plate of the fluid chamber was a cover glass that was cleaned by Piranha solution (30\% $H_{2}O_2$ and 70\% $H_{2}SO_4$). The distance between the cover glass and substrate surface was 300 $\mu$m. The length and width of the channel was 4.5 cm and 1.3 cm, respectively.

Solution A was a solution containing water, 1-octanol, ethanol (95\%, Fisher Scientific, Canada), and halochromic chemical compounds (BP or BG). The ratio of ethanol, water and 1-octanol in solution A was 60:26:14 by volume.  Colored droplets were formed by solvent exchange using solution A containing BG or BP. When BP was used as the indicator in the droplets, 0.1g of BP and 0.1g of sodium acetate were added into 50 ml solution A. Here sodium acetate (buffer, 99\%, Fisher Scientific, Canada) was used to adjust the initial pH of the droplets. When BG was the indicator in the droplets,  0.1g of BG was dissolved in 50 ml solution A.

Solution B was 1-octanol saturated by water. During the process of solvent exchange, the fluid chamber was filled with solution A that was then displaced by solution B at a controlled flow rate. Water droplets containing halochromic compounds were formed on the surface of the substrate after solvent exchange, as shown in Figure \ref{f1}b.

\subsection{Detection of acids in oil phase}
 After formation of the surface droplets, solution C was injected into the fluid cell at constant flow rate. Here solution C refers to water-saturated 1-octanol containing different types and concentrations of organic acids (i.e. target analytes). Solution C was made by directly dissolving the different acids into water-saturated 1-octanol by molar concentration. An upright optical microscope (Nikon H600l) equipped with 10X lens and a camera was used to capture the process of the color change of the droplets when the acids partitioned into the droplets, as shown in Figure \ref{f1}b and c. To quantitatively measure the concentration of the partitioned acid in droplets, the time required for droplet color change was recorded at different concentrations of the acid. The time of discolor refers to the time that the color intensity of the droplet reached a fixed value from contacting with the acid in the oil. The color intensity, known as the color luminance, was directly measured by ImageJ.

\subsection{Detection of acids in an organic phase by bulk mixing}

To compare with the droplet-based approach, we performed bulk solution-based detection of acids in an oil phase by directly mixing the oil with an aqueous solution containing halochromic compounds. A cosolvent ethanol was used to facilitate the mixing process. The halochromic compound, BP or BG, was dissolved in water-ethanol solution (water: ethanol = 60:26 by volume ratio) at the concentration of 0.04\% by mass ratio. Then 1-octanol containing acetic acid at different concentrations was added into the solution of the halochromic compound at 1:1 volume ratio, leading to formation of emulsion due to ouzo effect (oversaturation) \cite{Vitale200317,Zemb201616}. In another bulk mixing method, BP or BG was dissolved in water at the concentration of 0.04\% by mass, and one volume of the prepared solution was added into a half volume of 1-octanol containing acetic acid at different concentrations.

\subsection{Colorimetric identification of acid mixtures and anti-counterfeiting of alcohol spirits by droplets}

Acid mixtures were prepared by directly dissolving three different acids (acetic acid, gallic acid and benzoic acid) into 1-octanol with three different concentrations, as shown in Table \ref{mix1}. Those ratios of acid mixtures have the same pH value in water. The pH values of the acid mixtures were measured by pH meter (Fisher, USA).

Four popular Chinese Liquors, Erguotou, Fenjiu, Fenjiu(30 years hoard time) and Maotai, were purchased from a local liquor store. To extract the acids from the liquor, 200 $\mu L$ of each spirit sample was added into 10 mL 1-octanol. The mixture was shaken for a few seconds and left standing for overnight. Afterward the mixture separated into two phases. The top layer of octanol-rich phase was taken out (namely solution C) and  injected into the flow chamber to interact with the indicating droplets for acid detection. The pH value of each spirit was shown in Table \ref{pHjiu}. All the pH values were measured by a pH meter(Fisher, USA).
All the above experimental procedures were conducted at $25^{\circ}C$.

\section{Results and Discussion}

\subsection{Formation of colored droplets for acid extraction and detection}

\begin{figure}[htp]
	\includegraphics[trim={0cm 0cm 0cm 0cm}, clip, width=0.95\columnwidth]{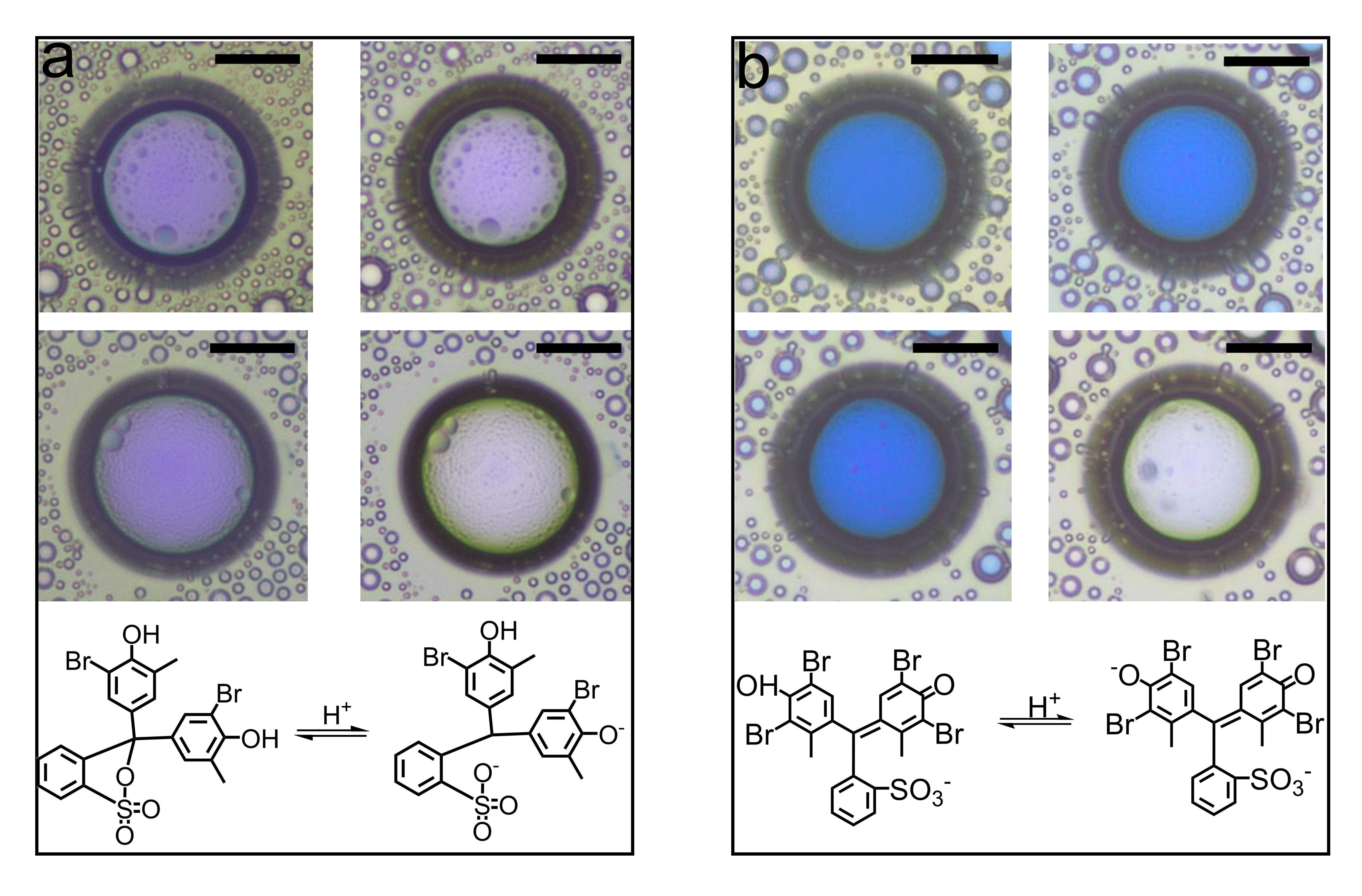}
	\caption{ Color of a droplet containing halochromic compound before and after contact with acid. Molecular structures of BP and BG are shown on the bottom row.  Top row: a droplet before (left) and after (right) in contact with 1-octanol for 5 minutes. Middle row: a  droplet before (left) and after (right) contact with octanol containing $1.7\times 10^{-4}M$ acetic acid for BP droplet in (a) and $1.0\times 10^{-4}M$ acetic acid for BG droplet in (b). Length of the scale bar: 25 $\mu m$.}	
	\label{f2}
\end{figure}

Following the process illustrated in Figure \ref{f1}, droplets prepared by the solvent exchange\cite{snd3} are few to tens $\mu$m in the lateral radius($R$). For droplet sizes in this range, the color change of the droplets during the colorimetric reaction can be clearly detected by optical microscope. At the same time, the polydispersed droplets enable us to examine the effect from droplet size.
These droplets contain aqueous solution of halochromic compounds, exhibiting different colors to indicate the pH level inside droplets. The color of droplets containing 
 two halochromic  compounds used in this work, bromocresol green (BG) and bromocresol purple (BP), is shown in Figure \ref{f2}. In an aqueous solution, the color of BG  changes from blue to light yellow as pH decreases from 5.4 to 3.8 while BP changes from dark purple to light yellow as pH decreases from 6.8 to 5.2. 
 
 We have added sodium acetate into solution A to adjust the pH value above 7 in the formed BP droplets. The initial color of the droplets was purple, as shown in Figure \ref{f2}a. Similarly, the initial color of BG droplets without sodium acetate was blue (Figure \ref{f2}b). The formation of colored droplets demonstrates that incorporating halochromic compounds into droplets can be achieved by the simple process of solvent exchange.

The optical images in Figure \ref{f2} show the color change of droplet in contact with the flow of oil containing acetic acid after 5 minutes. The extracted acid from oil reacts with the halochromic compound in the droplets, leading to the loss of droplet color. The indicating formula of BP and BG with acid are presented in Figure \ref{f2}. As control, the droplet color remained intact after the same period of time injection of pure 1-octanol.

\subsection{Sensitivity of colorimetric reaction in droplets}
The color of droplets is sensitive to the presence of acid in the oil. Figure \ref{f2}a shows that BP droplets changed color after the injection of oil containing acetic acid at a concentration of 0.17 $mM$, suggesting such acid concentration in the droplets was high enough to shift the equilibrium of BP into colorless product. At an even lower concentration of the acid (0.1 $mM$) in oil, BG droplets changed from blue to colorless. Below these concentrations of acetic acid in oil, the amount of the extracted acid in the droplets is so low that the droplets remain the original color. We simply refer this concentration of the acid in oil as the limit of detection (LoD) of the colorimetric reaction in droplets. The LoD of BP droplets for acetic acid in oil was around 0.17 $mM$ and of BG droplets were even lower, around 0.1 $mM$. 

We compared the LoD of droplets and two common shake-flask bulk methods shown in Figure S1 in SI. When the acid in oil was directly mixed with an aqueous solution of BP or BG, the dark blue color of the aqueous solution after oil-water phase separation changed to bright orange when the initial concentration of acetic acid in oil solution was above a certain level. The LoD was found to be 0.17 M for BP and 0.1 M for BG, three orders of magnitude higher than that of droplets. By the second method, acetic acid in oil was mixed with the solution of BG or BP in an ethanol aqueous solution (water: ethanol=60:26 by volume). Upon mixing, an emulsion formed due to the presence of the cosolvent ethanol through the well-known Ouzo effect \cite{Vitale200317}. The LoD was determined to around 0.17 M for BP mixture and is 0.01 M for BG mixture. Although the emulsion formation improves the sensitivity of the shake-flask bulk method, the LoD is still 2 orders of magnitude higher than that of droplets containing the same halochromic compound.

We attribute such sensitive response of colorimetric reactions in droplets to the preferable microcompartmentalized environment in the droplets. The partition of the acid from oil phase to water phase gives rise to a higher concentration of a water soluble acid in droplets. The limitation of droplet-based method is that acid has to be extracted into the droplets to react with the pH indicator in droplets, so the droplet color is less sensitive for those acids with long hydrophobic chains and much higher solubility in oil than in water. This droplet-based acid detection in oil is performed in a flow condition so it is fast, requiring a small amount of samples. Nontoxic solvent is involved in the entire process.

\subsection{Acid specificity of colorimetric reaction rate in droplets} 

\begin{figure}[htp]
	\includegraphics[trim={0cm 0cm 0cm 0cm}, clip, width=0.98\columnwidth]{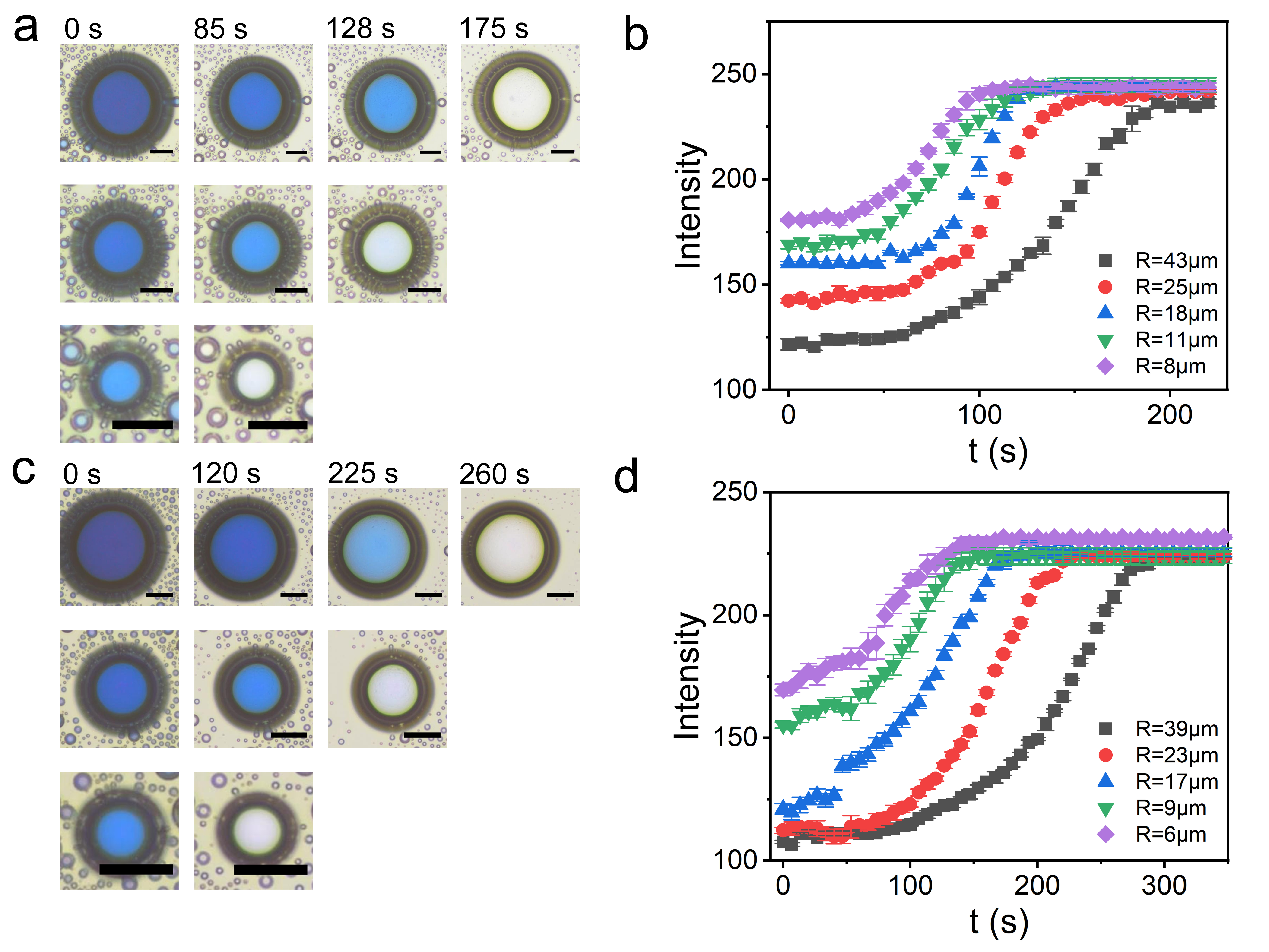}
	\caption{Detection of acetic acid in oil by surface droplets containing BG. (a) Time series images of droplet discoloration from $5\times 10^{-4}M$ acid in oil. The droplet size decreases from the top to the bottom row. Length of the scale bar :25 $\mu m$. (b) The color intensity at different  time after the droplets in contact with oil containing $5\times 10^{-4}M$. (c) Time series images of droplet discoloration from $1\times 10^{-4}M$ acid in oil. (d) The color intensity at different time after the droplets in contact with oil containing $1\times 10^{-4}M$. The error bar is from three repeats of the measurement.}
	\label{f3}
\end{figure}

Apart from high sensitivity, an even more powerful feature of droplet-based detection is time-dependent response of droplet color. 
Figure \ref{f3}a,c show the droplets of three representative sizes in contact with oil containing acetic acid of two different concentrations, 0.5 $mM$(a) and 0.1 $mM$(c). The plots showing the change of color intensity as a function of time within different sizes of droplets are presented in Figure \ref{f3}b,d, respectively. For 0.5 $mM$ of acetic acid, the required time ($t$) for droplet decolor (i.e., becoming colorless) was $\sim$ 75 s for a droplet diameter $R$ of 8 $\mu m$, and extended to $\sim$ 185 s for $R$ of 43 $\mu m$ (Figure \ref{f3}b). At given concentration of a specific acid, clearly the decolor time $t$ increases with the increase in the droplet size. The faster color change suggests that the colorless product reaches a critical concentration faster in smaller droplets. When the concentration of acetic acid was reduced to 0.1 $mM$, $t$ was 120 s for $R$ of 9 $\mu m$ and extended to 259 s for $R$ of 39 $\mu m$ (Figure \ref{f3}d). The results suggest that for lower concentration of acid, the decolor time of droplets is longer. In addition, the dependence of the decolor time on the droplet size becomes increasingly pronounced at a lower concentration of acid. 

\begin{figure}[htp]
	\includegraphics[trim={0cm 0cm 0cm 0cm}, clip, width=1\columnwidth]{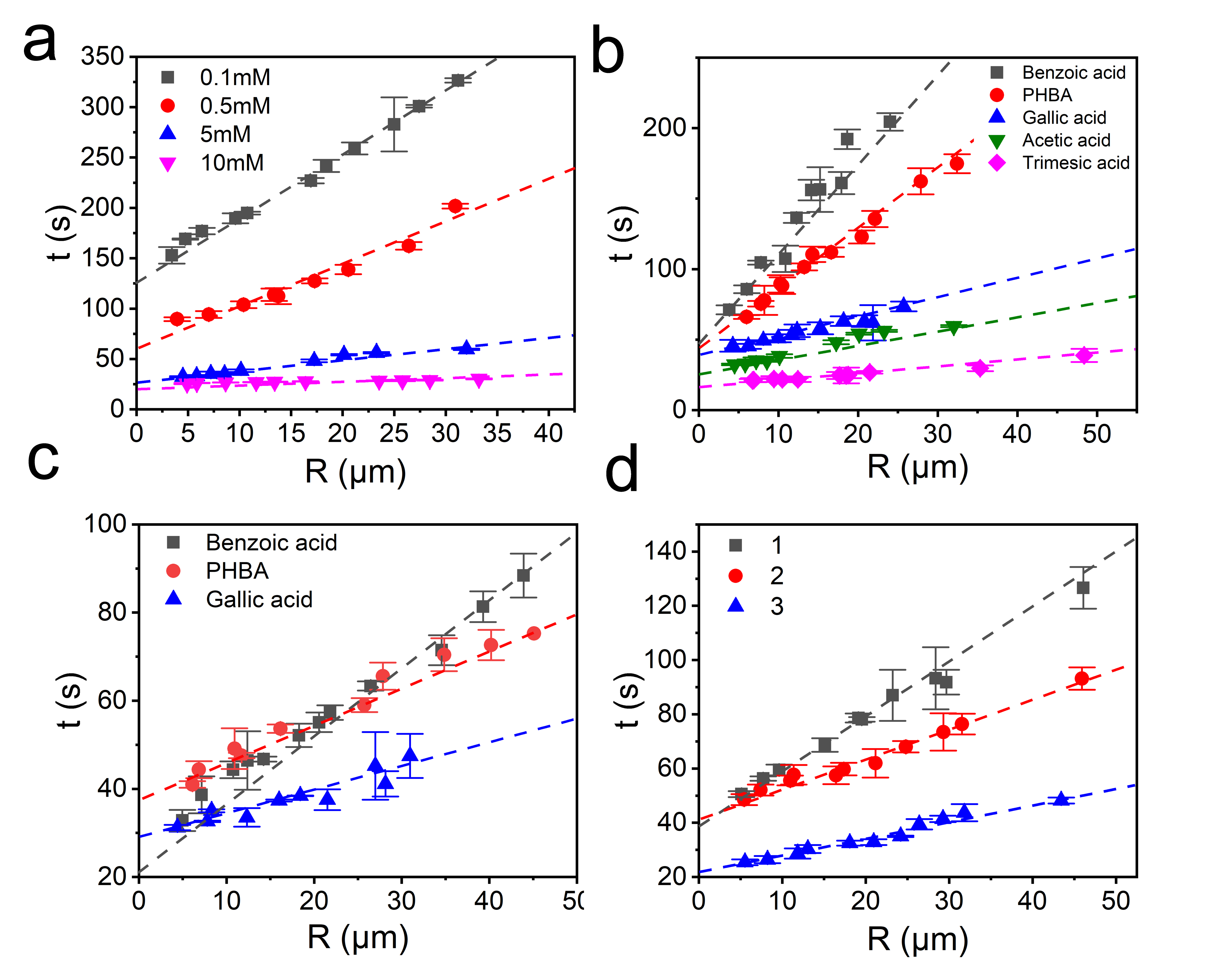}
	\caption{Detection of different acids and acid mixtures in oil by surface droplets containing BG. Time of the droplet discolor as a function of the droplet radius at different concentrations of acetic acid (a). The acid concentration is 5 $mM$ in (b) and 50 $mM$ in (c). (D) Detection of acid mixtures at different ratio in oil by surface droplets containing BG.  Acetic acid, gallic acid and benzoic acid were mixed at different ratio, as shown in Table \ref{mix1}. Dotted lines are the linear fitting. The error bar is from two to three repeats of the measurement.}
	\label{f4}
\end{figure}

The impact of droplet size on the colorimetric processing time was then systematically studied. Figure \ref{f4}a shows the plot of the decolor time $t$ against the droplet radius at different acid concentrations. Results reveal the dependence of decolor time on the droplet size is pronounced for larger droplets than smaller droplets. When the droplet radius is below 1 $\mu$m, there is not much difference in decolor time, possibly due to similar transport rate of the acid into such small droplets and the limitation of the instrument to identify the slight difference. Moreover, the decolor time is dependent on the acid concentration. That is, at lower acid concentration, the dependence of decolor time on the droplet size is more pronounced. In addition to acetic acid, we have also tested six other acids listed in Table \ref{lgP}. The results in Figure \ref{f4}b show similar trends for all acids at the concentration of 5 $mM$. Comparing with a higher concentration of 50 $mM$ (Figure \ref{f4}c), the change of time is sharper at lower acid concentration.

Most importantly, the decolor time of droplets $t$ exhibits specificity to the acid at a low concentration, that is, the decolor time for the same-sized droplets depends on the acid type. For instance, decolor time for a droplet of 20 $\mu$m in radius is 200 s in response to 5 $mM$ benzoic acid in the solution, but is only 25 s to trimesic acid at the same concentration. With shorter decolor time for smaller droplets for all acids, the decolor time is around 100 s for benzoic acid and 20 s for trimesic acid. There is also a significant difference between benzoic acid and gallic acid at 5 $\mu$m. As the acid concentration increases, the decolor time becomes so short that no clear difference can be measured for different types of acids. For instance at 50 $\mu$m, the decolor time of the droplet at any size is close to 20 s for trimesic acid and the same-sized droplet is similar for both gallic acid and benzoic acid.

To explain the acid specificity in decolor time of droplets, we  
 consider a simplified case where the transport of acid molecules from the oil into the droplet is a rate-determining step. Immediately after the intake into the droplets, acid molecules dissociate and react with pH indicator. The flow rate of the acid solution is constant in all measurements, so any influence of the external flow on the acid transport is expected to be the same for all experiments. The intake of the acid from the oil into the aqueous droplet is driven by the partition of acid in water. With time, the concentration of acid in water $C$ with time $t$ can be described by first-order reaction, as established in the early work on solvent extraction of a single droplet \cite{DDF1,DDF3,DDF4,DDF5}. 

\begin{equation}
\frac{dC}{dt}=\frac{A}{V}D(pC_o-C_w)
\label{6}
\end{equation}
Here the effective diffusion coefficient ($D$) of the acid in the oil is on the same order of magnitude for different acids, considering that the effect from Taylor-Aris dispersion is same for all experiments at the same flow rate\cite{tayer1,tayer2,tayer3,tayer4} and the diffusion coefficients of all acids are similar in the oil solution. $A$ is the surface area of the oil-water interface of the droplet where the intake of the acid takes place and $V$ is the volume of the droplet. $C_o$ and $C_w$ are the concentrations of the acid in the oil phase and the water droplet at time $t$, respectively. $p$ is the distribution coefficient $p = \frac{C_o}{C_w}$ at equilibrium state. At decolor time $t$, $C$ reaches $C^*$(the critical concentration) so that the proton from acid dissociation reaches the level that is sufficiently high for the color change of the pH indicator. After integration and transformation\cite{DDF1}, Eq.\ref{6} can be re-written.
\begin{equation}
t \sim \frac{R}{D}\ln (1-\frac{pC^*}{C_o})
\label{7}
\end{equation}

Where $R$ is the radius of the droplets. Based on Eq. \ref{7}, it is clear that from the same type of acid at given acid concentration, the decolor time is longer for larger droplets and the time increases linearly with increase in the size of the droplets. For different acids, the decolor time depends on the distribution coefficient p, which influences $C_w$ in the droplets. Both distribution coefficient and dissociation constant of the acid in the droplet contribute to the transfer rate of the acid into the droplets, giving rise to the acid specificity of the decolor time.

The timescale for the acid transport analyzed above is only one of the steps that can influence the decolor time. Other possible steps are dissociation of the acid to produce protons and eventually reaction kinetics of the acid with the pH indicator in the droplet.
For a weak acid with a long hydrophobic chain, the adsorption and desorption on the droplet surface may also play an important role in the decolor time. Future work is required to establish quantitative correlation between decolor time and all steps that contribute to the acid specificity.

Here we can also provide a good correlation between acid properties and their LoD in droplet. Considering the volume of droplets is very small, on an order of femtoliter, we can assume that the acid concentration in the droplets can reach an equilibrium with the oil phase. We estimate the concentration of $H^+$ in the droplets from the concentration in oil, partition coefficient, the dissociation coefficient in water. For a polyprotic acid, such as citric acid, we only consider the first order of dissociation because the second and third dissociation are very weak (with each decreasing in the orders of magnitude). The partition coefficient lgP is defined by the equilibrium constant of the acid in water and oil.

\begin{equation}
Lg P=Lg \frac{[HA]_{O}}{[HA]_{W}}
\label{P}
\end{equation}

\begin{equation}
[H^+]_{W} = (K \times [HA]_{W})^{1/2}
\label{H}
\end{equation}

$[HA]_O$ and $[HA]_W$ are the concentration of the acid in oil and the un-dissociated acid in water \cite{logp}, respectively. $K$ is the dissociation constant of the acid in water droplets.  

To show the dependence of LoD on lgP in oil and water, we determined the LoD of four organic acids with the partition coefficient ranging from -1.6 to 2.48. 
As listed in Table \ref{t2}, a lower partition coefficient of the acid in oil leads to a lower LoD and more sensitive detection by colorimetric reaction in droplets. Considering the same dissociation constant, the acid concentration in the droplets is expected to be higher for an acid with a higher partition coefficient at a given acid concentration in the oil phase.

 The correlation among partition coefficient LgP, dissociation  constant  and  LoD  of  acid in colorimetric reactions allows us to determine one parameter if the other two are known. For acetic acid, our measurements show that the minimum concentration in oil needed for the color change of droplets (i.e. LoD of acetic acid in oil) is $1\times 10^{-4}M$.  At the LoD, the corresponding concentration of acetic acid in water droplets $[HA]_W$ is determined to be $1.48\times 10^{-4}M$ according to Eq.\ref{P} with LgP of -0.17 based on the literature\cite{gallicacid2}. The dissociation constant (pKa) of acetic acid in water is 4.76 in literature \cite{citricacid1}. Therefore for $1.48\times 10^{-4}M$ of $HA$ in droplets, the concentration of $[H^+]$ from dissociation is $5.07\times 10^{-5}M$ (pH of 4.29), calculated from Eq.\ref{P} and \ref{H}.  This calculated  $[H^+]$ in the droplets is in good agreement with the range of pH level for BG to change from yellow to blue at pH of 3.8 to 5.4. Similarly, if we know both the dissociation constant of the acid in water and the partition coefficient of the acid between water and oil, we can predict the LoD of the droplets in our measurements, which is the minimum concentration of a certain acid in the oil phase in equilibrium with the concentration of $H^+$ in the droplets for color change. Identically, for a given pH indicator($H^+$ is known), when we know the LgP and pKa from the literature, the theoretical LoD($HA_O$) can be calculated by using Eq. 3 and 4. The calculated LoD of acids are shown in Table \ref{t2}. The calculated LoD of all acids, except citric acid, have good agreement with the experimental one, suggesting the important role of participation coefficient and dissociation constant in the colorimetric reactions in the droplets. The calculated LoD  of citric acid is very low due to its low LgP. At very low concentration of citric acid, much longer time is needed for the completion of extraction process, as revealed by Eq. 2. In our experiments, the time window for decoloration is around 5 minutes to ensure the droplets cannot be dissolved by the continuous solution flow. This time window may not be sufficient for complete extraction of citric acid, which may lead to higher LoD from the experiment than that from calculation.
 Although the dependence of decolor time on the type of acid is not as specific as fingerprints in molecular spectra, this specificity can be very useful when we compare acid matrix in a solution. We demonstrate this concept by using model mixtures of acids and then apply the concept for identifying alcoholic drinks of good and bad quality in anti-counterfeiting. Figure \ref{f4}d shows the detection of acid mixtures with compositions listed in Table \ref{mix1}. With three different combinations of three acids, the pH value of all three mixtures is at 3.36 as determined from the measurements by using a pH meter. However, when BG droplets are in contact with oil containing the acid mixture, the decolor time $t$ is different especially for larger droplets. It takes the longest for mixture 1 and the shortest for mixture 3. Therefore the decolor time of the droplets can be used for identifying whether the acid profile is the same when the total acid level is the same.

\subsection{Colorimetric analysis of acid mixtures for anti-counterfeiting alcoholic drinks}

To demonstrate one possible application based on colorimetric reactions in droplets, we show an identification of a counterfeit alcohol spirit by evaluating the profile of organic acids in the alcoholic beverage.  Counterfeit spirits are normally inferior spirits to replace upmarket spirits.  For instance, a typical counterfeit Maotai may be a cheap spirit on the market or even a home-made one. Our droplet-based method is capable of distinguishing different types of spirits because each spirit has its characteristic acid profile. We expect such a complex acid matrix can give distinct decolor time for each spirit.

\begin{figure}[htp]
	\includegraphics[trim={0cm 0cm 0cm 0cm}, clip, width=0.95\columnwidth]{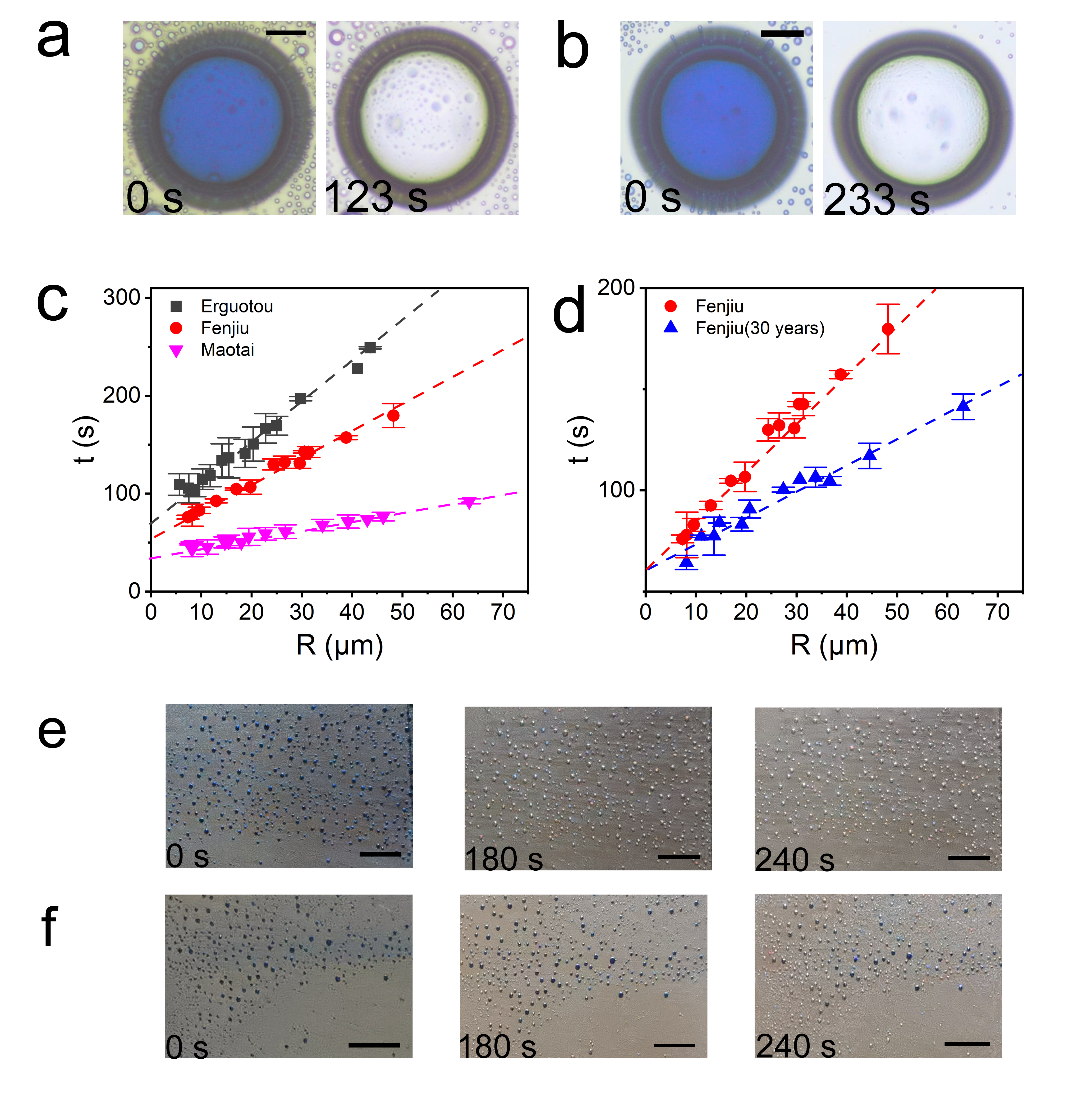}
	\caption{Determination of counterfeit Chinese spirit through surface droplet based nanodetector and the equipment-free detection. (a) Time series pictures of the droplet discoloration at Fenjiu. (b) Time series pictures of the droplet discoloration at Erguotou. (c) Time of the droplet discolor VS the reciprocal of the droplet size at different spirit.(d) Time of the droplet discolor VS the reciprocal of the droplet size at two types of Fenjiu. Dotted lines are the linear fitting. The error bar is from three repeats of the measurement. The images of the droplet decoloration in contact with oil that contains compounds extracted from  Fenjiu (e) and from Erguotou (f) at 0 s, 180 s and 240 s by using a commercial smartphone camera. (Length of scale bar: 25 $\mu m$ in a,b and 2 $mm$ in e and f)}
	\label{f6}
\end{figure}

We tested four most popular Chinese spirits, Erguotou, Fenjiu (less expensive), Fenjiu (30 years) and Maotai. Erguotou and Fenjiu are the same type of Fen-flavor, while Maotai is soy sauce flavor. All of these four spirits contain over 27 types of organic acids from the brewing process with a high concentration (over 735 mg/L)\cite{spirit1,spirit2}. Figure \ref{f6}a and b show the time series images of the droplets changing color in contact with oil containing the organic acids  extracted from Fenjiu and Erguotou. At a similar droplet size, it takes 123 s for the droplets to decolor by interacting with Fenjiu, while a substantially longer time of 233 s, is required for the Erguotou samples prepared in the same way. The difference of $t$ for Fenjiu and Erguotou samples demonstrates the potential application to distinguish different spirits by our method. Figure \ref{f6}c shows decolor time $t$ versus the droplet size $R$ for Erguotou, Fenjiu, and Maotai. The time for the droplets to complete color change is longer for Erguotou than other spirits, indicating that the acid profiles of these two spirits are different, even they have a similar pH value. 
A higher ratio of the hydrophilic acids in the sprits will lead to faster decoloration of the droplets. For Fen-flavor spirit, Fenjiu (30 years) is the most expensive and the cheapest one is Erguotou. Our method also can distinguish the spirit with different hoard time in the same brand (same feed stock and process). Figure \ref{f6}d shows the time for the droplets to complete color change is shorter for Fenjiu (30 years) than common Fenjiu. Long hoard time of the spirit would help the long chain organic acid to break into short chain organic acid and help development of richer, more complex and more pleasant organic acid profile.
The above results demonstrate that our surface droplet-based colorimetric detection can provide a fast reference to evaluate the organic acids profile, which is effective in distinguishing authentic spirit from the counterfeit samples.

Finally, we note that the droplet-based spirit detection can be possibly acquired by simply using the camera on a smart phone. 
Figure \ref{f6}e and f show the images of the colored droplets with visually observed sizes are captured by a smartphone camera Huaiwei(P30 Pro). When the sample extracted from the Fenjiu was introduced into the detector, a distinct decolor of the droplets was observed within 180 s, as presented in Figure \ref{f6}e. In contrast, the decolor of the droplets can not be completed within the same period when the detection of Erguotou was carried out, as presented in Figure \ref{f6}f. The fast and optical assessment to the acid level in beverages may be potentially applied to distinguish the spirits with different years of aging (e.g., 5 years versus 10 years), based on the difference in the acid profile in the drinks. The results demonstrate the low-cost and equipment-free potentials of the colorimetric detection based on surface droplets for broad applications.

\section{Conclusion}

In summary, we have demonstrated a novel approach for combinative analysis of mixtures based on functional surface nanodroplets. The model analysts are organic acid mixtures. The halochromic compound in aqueous nanodroplets changes color in response to the dissociation of extracted acids from the oil. The coupled effect of partition coefficient and dissociation constant determines the nanoextraction rate, giving rise to the specificity of decolor time on the acid type. 
Such specificity of the colorimetric reactions in the droplets provides a sensitive method to compare the acid profiles in multiple component mixtures without time-consuming and expensive equipment. We demonstrate the application of nanodroplet-based combinative analysis in identifying counterfeit spirits of different brands. The principle behind the combinative analysis is general, relying on nanoextraction and chemical reactions in femtoliter droplets. The detection is obviously not limited to simple organic acids in our model systems, but applicable to many compounds that are extractable from sample solutions. 


\section{Acknowledge}
M.L. acknowledges the support from Vice-Chancellor’s Postdoctoral Fellowship from RMIT. The project is supported by the Natural Science and Engineering Research Council of Canada (NSERC) and Future Energy Systems (Canada First Research
Excellence Fund). This research was undertaken, in part, thanks to funding from the Canada Research Chairs program. We thank Qiuyun Lu for valuable discussion on partition coefficient analysis and Jae Bem You for the comments on the theoretical analysis.

\newpage
\begin{table}[ht]
 \centering
 
 \caption{Molar mass, pKa and LgP of each acids from literature}
 
\begin{tabular}{p{4.5cm}|p{4cm}|p{2cm}|p{3cm}}
\hline
Acids& Molar mass /$g\times mol^{-1}$ &pka\cite{citricacid1,Acetic,heptanoic,Gallic,PHBA,BA,trimesic1}&  LgP \cite{citricacid2,Acetic,heptanoic,Gallic,PHBA,BA,trimesic2}\\
\hline
Citric acid& 192.12  & $3.13_{pKa1}$&  -1.64\\
Acetic acid& 60.05  & 4.76& -0.17\\
Heptanoic acid& 130.18  & 4.4 & 2.42\\
Gallic acid& 170.12  & 4.4& 0.7\\
4-Hydroxybenzoic acid& 138.12  & 4.54& 1.58\\
Benzoic acid& 122.12  & 4.2& 1.87\\
Trimesic acid& 210.14  & $3.12_{pKa1}$& -0.56\\

\hline
\end{tabular}
\label{lgP}
\end{table}
\begin{table}[ht]
 \centering
 
 \caption{The contents of different acid mixtures.}
 
\begin{tabular}{p{3cm}|p{3cm}|p{3cm}|p{3cm}|p{2cm}}
\hline
Acid mixtures& acetic acid/M& gallic acid/M& benzoic acid/M& pH\\
\hline
1& 2.6$\times10^{-4}$&1.7$\times10^{-3}$&2.0$\times10^{-4}$&3.36\\
2& 1.8$\times10^{-5}$&3.3$\times10^{-4}$&3.0$\times10^{-3}$&3.36\\
3& 5.6$\times10^{-3}$&1.6$\times10^{-4}$&1.2$\times10^{-4}$&3.36\\

\hline
\end{tabular}
\label{mix1}
\end{table}
\begin{table}[ht]
 \centering
 
 \caption{pH value of each Chinese spirit}
 
\begin{tabular}{p{2cm}|p{3cm}|p{3cm}|p{3cm}|p{3cm}}
\hline
Spirits& Erguotou & Fenjiu& Fenjiu(30 years)& Maotai\\
\hline
pH& 3.76&3.49&3.47&3.46 \\
\hline
\end{tabular}
\label{pHjiu}
\end{table}
\begin{table}[ht]
 \centering
 
 \caption{LOD of different acids by using BG droplets and the calculated LgP of each acids.}
 
\begin{tabular}{p{3cm}|p{3.5cm}|p{3.5cm}|p{4cm}}
\hline
Acids& LoD from experiment & LgP from calculation& LoD from calculation\\ 
\hline
Citric acid& 0.001$mM$& -0.55& 8$\times$ $10^{-5}$ $mM$\\
Acetic acid& 0.1$mM$& -0.17& 0.1$mM$\\
Gallic acid& 0.5$mM$& 0.88& 0.32$mM$\\
Heptanoic acid& 50$mM$ & 2.48& 16$mM$\\

\hline
\end{tabular}
\label{t2}
\end{table}
\newpage
\bibliography{nanoextraction}
\newpage

\end{document}